# Optical Vortex Dynamics in non-uniform twisted Ring-Core Fibers

**HASSAN ASGHARZADEH B.,** * **REGINA GUMENYUK, AND MARCO ORNIGOTTI**

*1Tampere University, Photonics Laboratory, Physics Unit, Tampere FI-33720, Finland*
*\*hassan.asgharzadehbastehdimi@tuni.fi*

**Abstract:** In this work, we present a thorough analysis of the propagation of fiber modes carrying orbital angular momentum in twisted, tapered, ring-core optical fibers. In particular, by generalizing the usual coupled-mode approach to include the effect of twisting and tapering, we discuss how it is possible to achieve efficient power transfer between modes carrying different amounts of orbital angular momentum. Our simulation allows us to get a clear insight into the dynamics of vortex modes propagating through twisted ring core fibers.

## 1. Introduction

The concept of beams carrying orbital angular momentum (OAM) was first introduced by Allen et al. [1], who demonstrated that light beams with a helical phase structure carry OAM. Since then, optical vortices (OVs), have drawn significant attention due to their applications in optical communication [2], imaging [3], and optical tweezers [3,4].

OAM beams can be generated either by free-space systems [3], or in optical fiber-based systems [5]. While the former often faces challenges like insertion losses, the latter are more suitable for high-power applications [6]. Various techniques have been developed for OAM beam generation in fibers, including tilted fiber Bragg gratings [7], long-period fiber gratings [6,8,9], mode-selective couplers [10], and helically twisted photonic crystal fibers [11,12]. Among these, twisted fibers are particularly useful, as the applied twist establishes OAM modes as the preferred basis for the guided modes [13].

Stable transmission of OAM beams is critical for practical applications. Alexeyev et al. [14] demonstrated that conventional fibers are not the best means for stable OAM beam transmission due to modal crosstalk effects. In contrast, ring-core fibers (RCFs) minimize modal crosstalk, making them better candidates for long-distance transport [5]. In addition, passive helical fiber gratings and twisted structures have attracted significant interest as efficient platforms for generating higher-order OAM modes, enabling direct conversion of a Gaussian beam launched into a fiber into a specific OAM mode without additional passive elements [15–17].

Motivated by all these considerations, in this work we thoroughly investigate OAM dynamics in tapered, twisted, RCFs. To stablish a more general framework, we model the fiber as a non-uniform case, for which the uniform case can be readily obtained by neglecting the terms associated with nonuniformity. Moreover, tapered fibers are of particular interest due to their promising applications in high power fiber amplifiers and lasers. In particular, we analyze the design of twisted, tapered RCFs and demonstrate how to efficiently transfer power from the fundamental mode to high-order OAM modes by precisely controlling the twisting parameters. To achieve this goal, we employ a two-step approach. First, using the formalism of transformation optics (TO) [18], we perform full vectorial finite element modeling (FEM) in COMSOL Multiphysics® [19] to characterize the guided modes of twisted RCFs. Second, we derive generalized coupled mode equations (CMEs) that describe OAM mode evolution under twist- and taper-induced couplings. This framework provides a clear insight into the dynamics of OAM modes propagating through twisted RCFs.

This paper is organized as follows: in Sect. 2, we present a detailed theoretical investigation of modal interactions in twisted, tapered RCFs. Section 3 integrates all the derived equations into a comprehensive model for analyzing OAM mode dynamics in twisted, tapered RCFs. Finally, Conclusions are drawn in Sect. 4.

## 2. Theory and numerical analysis

### 2.1 Guided mode analysis of twisted ring-core fibers

The modal analysis of twisted RCFs is, in general, challenging, as twisting introduces a $z$-dependent refractive index, making it necessary to adopt a full 3D approach. However, TO [18,20,21], provides a powerful tool to overcome this issue by mapping the evolution of light in a twisted fiber into an equivalent straight fiber, where the effect of twisting is mapped into an anisotropic permittivity and permeability [21]. Based on this formalism, in fact, any coordinate transformation can be equivalently expressed as a modification of the permittivity and permeability of the medium. For fibers twisted along the propagation direction ($z$-axis), the mapping is achieved by a transformation from Cartesian laboratory coordinates $(x, y, z)$ to co-rotating helical coordinates $(x', y', z')$ as [20,21]

$$\begin{cases} x' = x \cos(\Upsilon z) - y \sin(\Upsilon z), \\ y' = x \sin(\Upsilon z) + y \cos(\Upsilon z), \\ z' = z, \end{cases}$$

(1)

where, $\Upsilon = \frac{2\pi}{\Lambda}$ [rad/mm] represents the twist rate and $\Lambda$ [mm] is the twist pitch. Throughout this work, we consider any quantity with a superscript (') as a quantity in the helical coordinate frame. Since Maxwell's equations are covariant under coordinate transformation, the electric field of the guided modes satisfies the same wave equation in both frames [21]

$$\nabla \times (\mu_r^{-1} \nabla \times \mathbf{E}) - k_0^2 \, \varepsilon_r \mathbf{E} = 0,$$

(2)

where $\mathbf{E}$ is the electric field vector, and $\varepsilon_r$ and $\mu_r$ represent the relative permittivity and permeability of the medium which, in general, can be anisotropic and position-dependent.

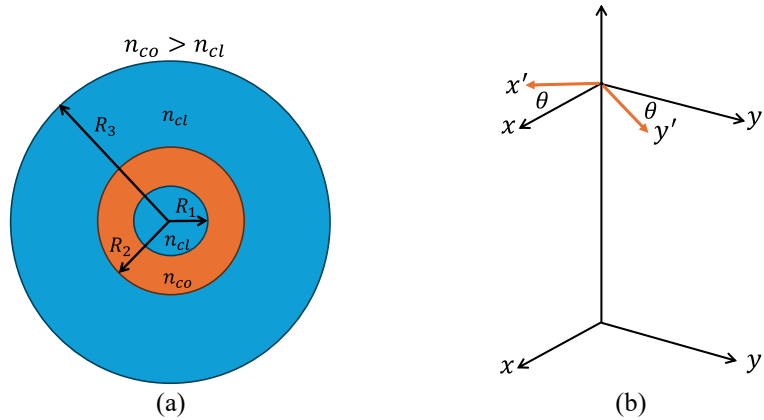

Fig. 1. (a) Cross sectional view of a typical RCF. The refractive indices of the core and cladding are $n_{co}$ and $n_{cl}$, respectively. The inner and outer radii of the core are represented by $R_1$ and $R_2$, respectively, while $R_3$ is the outer cladding radius. (b) The relation between Cartesian and helical coordinate frames. The laboratory frame is described by $\{x, y, z\}$, whereas the helical frame is denoted by $\{x', y', z'\}$. Parameter $\theta$ represents the twist angle.

In this study, we assume the original RCF to be an isotropic, non-magnetic, lossless, step-index medium with $\mu_r = 1$ and $\varepsilon_r = n^2(x, y)$. Here, $n$ is the original refractive index, with $n_{co}$ and $n_{cl}$ specifying its values in the core and cladding, respectively (See Fig. 1). Applying the coordinate transformation in Eq. (1) to Maxwell's equations yields the equivalent permittivity and permeability tensors in the helical frame [20]

$$\begin{cases} [\boldsymbol{\varepsilon}'_r(x',y',z)] = \varepsilon_r(x',y',z) \, [\boldsymbol{T}]^{-1}, \\ [\boldsymbol{\mu}'_r(x',y',z)] = \mu_r(x',y',z) \, [\boldsymbol{T}]^{-1}, \end{cases} \quad (3)$$

where

$$[\boldsymbol{T}]^{-1} = \begin{bmatrix} \Upsilon^2 y'^2 + 1 & -\Upsilon^2 x'y' & -\Upsilon y' \\ -\Upsilon^2 x'y' & \Upsilon^2 x'^2 + 1 & \Upsilon x' \\ -\Upsilon y' & \Upsilon x' & 1 \end{bmatrix}. \quad (4)$$

For simplicity of notation, we denote a tensor quantity X by [X]. Since $[\boldsymbol{T}]^{-1}$ is independent of $z'$, the modal field distributions in the helical frame depend on z (or equivalently $z'$ as $z = z'$) only through $e^{i\beta' z'}$. This translation invariance reduces the 3D problem of analyzing guided modes in twisted RCFs, in the laboratory frame, to a 2D problem in helical frame. In Fig. 2, we compare the intensity and phase distributions of the first few modes in an untwisted (upper panel) and twisted (lower panel) RCF, computed using COMSOL in the co-rotating frame. In the untwisted case ($\Upsilon = 0$ in Eq. (1)), the co-rotating and laboratory frames coincide. These results confirm that vector OAM modes are the most convenient basis for describing guided modes in twisted RCFs. Accordingly, we use the following expressions to express the electric field of circularly polarized OAM modes, $\boldsymbol{\xi}_l(r', \varphi')$ [22]:

$$\boldsymbol{\xi}_l(r', \varphi') \equiv \begin{cases} CV^{\pm}_{\pm l,m} = \{F_{l,m}(r')(\hat{r}' \pm i\hat{\varphi}') + F_{z,l,m}(r')\hat{z}\}e^{iJ_l\varphi'} & l \geq 0 \\ CW^{\mp}_{\pm l,m} = \{F_{l,m}(r')(\hat{r}' \mp i\hat{\varphi}') + F_{z,l,m}(r')\hat{z}\}e^{iJ_l\varphi'} & l > 1 \\ TM_{0,m} = F_{1,m}(r')\hat{r}' + F_{z,1,m}(r')\hat{z} & l = 1 \\ TE_{0,m} = -F_{1,m}(r')\hat{\varphi}' & l = 1 \end{cases} \quad (5)$$

where $l$ is the topological charge (OAM content) of the corresponding mode, and $(\hat{r}' \pm i\hat{\varphi}')$ correspond to left (+) and right (−) circular polarization states. The total angular momentum (TAM) of an OAM mode of order $l$ is given by $J_l = l + s$, where $s = +1 \, (-1)$ correspond to the spin angular momentum (SAM) of left- (right-) handed circular polarization. The terms $CV^{\pm}_{\pm l,m}$ and $CW^{\mp}_{\pm l,m}$ refer to SAM co-rotating and counter-rotating OAM modes, respectively. In the helical frame, the radial and azimuthal coordinates are given by $r' = \sqrt{(x')^2 + (y')^2}$ and $\varphi' = \tan^{-1}(y'/x')$, with $\hat{r}'$ and $\hat{\varphi}'$ as the corresponding cylindrical unit vectors. The function $F_{l,m}(r')$ denotes the radial profile of the transvers field, while $F_{z,l,m}(r')$ represents the radial distribution of the longitudinal component [23].

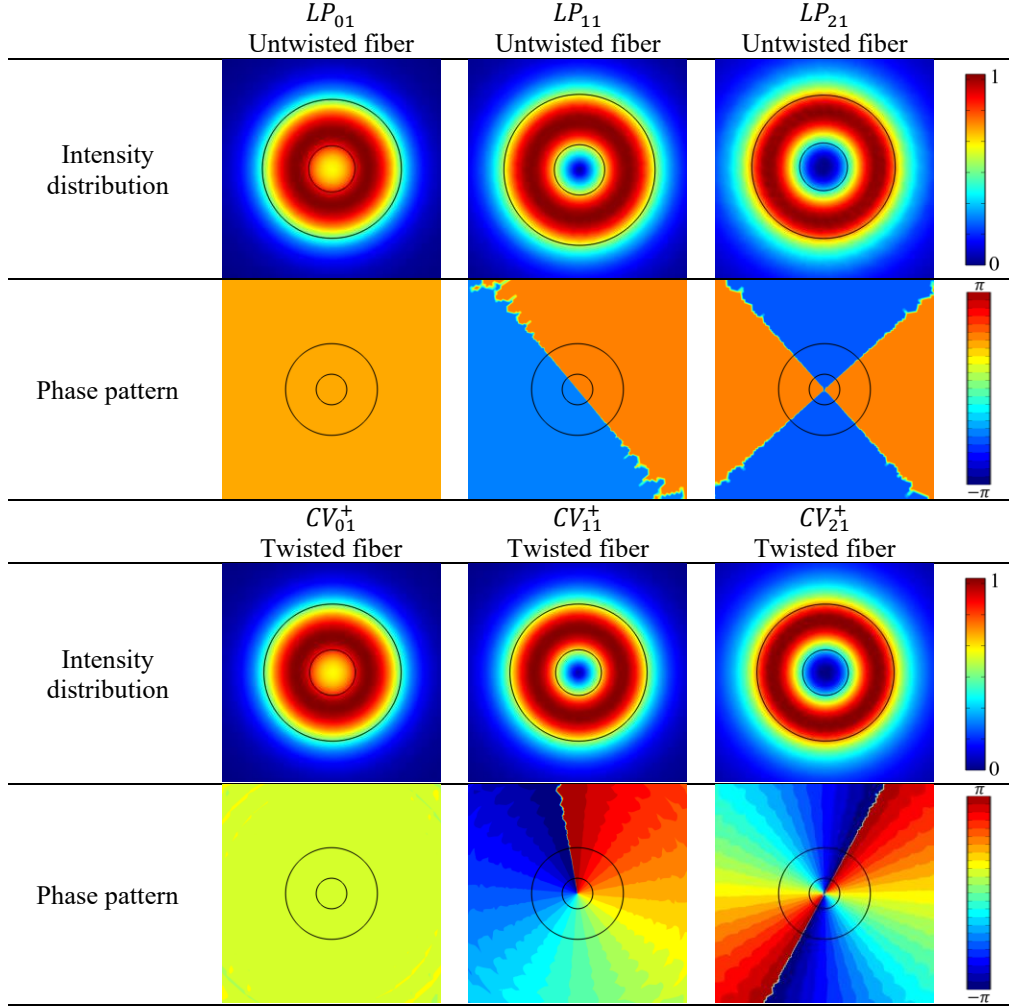

Fig. 2. Intensity and phase distributions of the first three eigenmodes of an RCF. Upper panel: Untwisted ($Y = 0$) RCF with modes $LP_{01}$, $LP_{11}$, and $LP_{21}$. Due to the ring-core geometry of the fiber, these untwisted modes exhibit a donut-shaped intensity distribution. For clarity, only the region surrounding the core is shown; the remaining outer cladding is omitted. Lower panel: The same modes in the same RCF with twist applied ($Y \neq 0$). Twisting the fiber transforms the untwisted modes into twisted ones ($LP_{11} \to CV_{11}^+$, and $LP_{21} \to CV_{21}^+$), as evident from their phase distributions, while the fundamental mode remains untwisted ($LP_{01} \to CV_{01}^+$). The spin, orbital and total angular momenta of $CV_{01}^+$, $CV_{11}^+$, and $CV_{21}^+$ modes are (+1, 0, +1 ), (+1, +1, +2), and (+1, +2, +3), respectively. Simulations were performed using COMSOL in the co-rotating frame.

## 2.2 Coupled mode theory

Having simplified the problem of guided mode analysis in twisted RCFs by mapping the twisted RCF into an equivalent straight one, we now establish a coupled mode framework to analyze modal dynamics. From the electromagnetic point of view, the twisted fiber in the laboratory frame and its counterpart, hereafter referred to as the *effective fiber*, in the co-rotating frame, are exactly interchangeable. This allows the modes of a straight fiber, with the same geometry and properties as the original one, to serve as a basis for describing light evolution in the twisted fiber. We can then compute the coupling coefficients using the equivalent permittivity and permeability of the effective fiber. To clarify this concept, we divide the analysis into two steps. First, we define an ideal RCF in the co-rotating frame, with permittivity and permeability matching those of the twisted fiber at $z = 0$, i.e.,

$$\varepsilon_r'(x', y', z' = 0) = \varepsilon_r(x, y) = n^2(x, y) \equiv \bar{\varepsilon}_r,$$

$$\mu_r'(x', y', z' = 0) = \mu_r(x, y) = 1 \equiv \bar{\mu}_r,$$

(6)

where the barred quantities refer to the ideal fiber. Using perturbation theory, the coupling coefficients are then calculated by treating the equivalent permittivity and permeability as perturbations to their ideal ones. Applying this approach first to tapering (Sect. 2.2.1) and then to twisting (Sect. 2.2.2), we derive a generalized set of CMEs that link mode evolution to both twist and taper parameters.

### 2.2.1 Taper-induced coupling coefficients

A local change in fiber radius can induce modal interaction during propagation. According to the coordinate transformation in Eq. (1), we note that $\sqrt{x^2 + y^2} = \sqrt{(x')^2 + (y')^2}$ at any $z$, implying $r = r'$. Therefore, any radial variation in the twisted fiber in the laboratory frame $(x, y, z)$ directly corresponds to the same variation in the effective fiber in the helical frame $(x', y', z)$. For a tapered step-index RCF, the refractive index depends on $z$ and reads

$$\left(n_{tap}(r', z)\right)^2 = n_{cl}^2 + NA^2\{H(r' - R_1(z)) - H(r' - R_2(z))\},$$

(7)

where $R_1(z)$ and $R_2(z)$ are $z$-dependent, inner and outer radii of the core, respectively (See Fig. 1). Here, $H(x)$ denotes the Heaviside step function [23], and $NA^2 = n_{co}^2 - n_{cl}^2$ represents the numerical aperture, with $n_{co}$ and $n_{cl}$ being the core and cladding refractive indices of the original fiber. A detailed derivation of the coupled local mode equations for non-uniform fibers is given in Ref. [23]. Since we follow the same approach, here we only present the final result. For tapered fibers, the local coupled mode equations for the $j$-th forward-propagating mode can be written as

$$\frac{da_j(z)}{dz} - i\beta_j(z)a_j(z) = \sum_q a_q(z)\,\Omega_{jq}(z),$$

(8)

where $\beta_j(z)$ and $a_j(z)$ are the propagation constant and modal amplitude, respectively, and $\Omega_{jq}(z)$ is the taper-induced coupling coefficient between modes $j$ and $q$, given explicitly as

$$\Omega_{jq}(z) = \left(\frac{c\,\varepsilon_0\,k_0}{4(\beta_j(z) - \beta_q(z))}\right) \iint \left\{\boldsymbol{\xi}_{jt}^*(x', y'; \beta_j(z)) \cdot \boldsymbol{\xi}_{qt}(x', y'; \beta_q(z))\right\} \frac{\partial (n_{tap}')^2}{\partial z} dx'\, dy'$$

(9)

with $t$ denoting the transverse part of the local electric field, $\boldsymbol{\xi}_{jt}(x', y'; \beta_j(z))$. The local mode fields of non-uniform fibers can be constructed by approximating the fiber as a series of individual segments of length $\delta z$, each treated as locally uniform [23]. Within each finite section, both $\boldsymbol{\xi}_{jt}(x', y')$ and $\beta_j$ can be considered $z$-independent. Accordingly, within each section $\Omega_{jq}(z)$ remains constant. Hence, Eq. (8) can be solved iteratively, using the output of each section as the input to the subsequent one.

Alternatively, we use an efficient numerical approach to solve Eq. (8). In this method, propagation constants and coupling coefficients are first computed at each segment along the fiber. These discrete values are then interpolated into smooth $z$-dependent functions, as illustrated in Fig. 3(a). Substituting these $z$-dependent functions into Eq. (8), the system can be solved numerically. When variations with respect to $z$ are sufficiently slow, this approach can reduce computational cost by decreasing the number of required discrete segments.

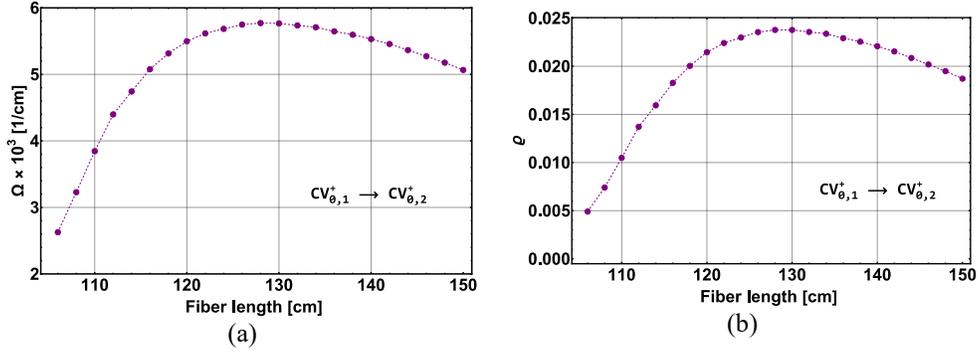

Fig. 3. (a) Coupling coefficient $\Omega$, between $CV_{0,1}^+$ and $CV_{0,2}^+$ due to tapering. (b) The ratio $\varrho$ for the corresponding mode coupling

We then substitute the refractive index of the ideal fiber from Eq. (6) into Eq. (7) and compute the coupling constants $\Omega_{jq}(z)$ using the OAM vector fields given in Eqs. (5). In doing so, we notice the emergence of a set of matching rules for propagation constants, spin and total angular momentum, expressed as (see Supplement 1)

$$\begin{cases} \beta_q \approx \beta_j & (a), \\ s_q = s_j & (b), \\ J_q = J_j & (c). \end{cases} \quad (10)$$

From Eq. (10), it follows that tapering permits coupling only between modes with the same polarization state and total angular momentum as well. Thus, $\Omega_{jq} = 0$ when $J_q \neq J_j$. Since only modes with the same polarization states can couple (see Supplement 1), the TAM selection rule in Eq. (10c) further requires them to share the same OAM content, i.e., $l_q = l_j$. Consequently, tapering allows coupling only to higher order radial modes.

These matching rules allow us to derive an analytical expression for the coupling coefficient $\Omega_{jq}$ (see Supplement 1), given by

$$\Omega_{jq}(z) = \left(\frac{\pi c \varepsilon_0 k_0}{\beta_j - \beta_q}\right) NA^2 \{R_2(z)\dot{f}_2(z) F_{j,m}(R_2(z)) F_{q,m}(R_2(z)) \\ - R_1(z)\dot{f}_1(z) F_{j,m}(R_1(z)) F_{q,m}(R_1(z))\}. \quad (11)$$

Fig. 3(a) illustrates the magnitude of the taper-induced coupling coefficient between the OAM modes $CV_{0,1}^+$ and $CV_{0,2}^+$, along the length of a tapered RCF, assuming that only $CV_{0,1}^+$ is initially excited at the narrower side. All the simulation parameters are listed in Table 1. This fiber supports modes with $l = 0$ and 1 at the narrower side of the taper, while at the wider side, it supports modes with $l = 0, 1, 2$ and 3. According to simulation data obtained from COMSOL, the fiber begins to support $CV_{0,2}^+$ at $z \approx 106\ cm$. Beyond this point, according to Eq. (10), the coupling occurs from $CV_{0,1}^+$ to $CV_{0,2}^+$. Given the coupling coefficient, $\Omega$, Eq. (8) can be solved numerically for all participating modes.

Numerical analysis reveals that taper-induced coupling is negligible as $CV_{0,1}^+$ retains almost all input power. Two primary factors may account for this behavior. First, the taper-induced coupling coefficients are numerically small, particularly when compared to the twist-induced coupling coefficients, as will be discussed in the next section. Second, the coupled modes are not phase-matched. According to coupled mode theory (CMT), in fact, efficient power transfer between two modes $p$ and $q$ occurs when the ratio $\varrho = 2|\Omega_{pq}|^2/(\beta_p - \beta_q) \gg 1$. As shown in

Fig. 3(b), $\varrho$ is much smaller than unity for the choice of parameters given in Table 1, confirming negligible power transfer between interacting modes.

| Table 1. Simulation parameters used in this study | | | |
|---|---|---|---|
| $n_{co}$ | 1.4543 | Inner radius of the core at the wider side of the taper, $R_1^w$ [$\mu m$] | $\Pi \times R_1^s$ |
| $n_{cl}$ | 1.4503 | Outer radius of the core at the wider side of the taper, $R_2^w$ [$\mu m$] | $\Pi \times R_2^s$ |
| $\lambda$ [$\mu m$] | 1.039 | Outer cladding radius at the wider side of the taper, $R_3^w$ [$\mu m$] | $\Pi \times R_3^s$ |
| Wider to smaller ratio, $\Pi$ | 2 (Fig. 3) | Tapering profile | $R(z) = \frac{b_0 - b_f}{2L} z^2 + \frac{b_f}{2} + 2R^s$ |
| | 2.5 (Fig. 4, Fig. 5) | | |
| | 1.64 (Fig. 6, Fig. 7) | | |
| Inner radius of the core at the narrower side of the taper, $R_1^s$ [$\mu m$] | 1.4167 | Average taper angle, $b_0$ | $2\frac{R^w - R^s}{L}$ |
| Outer radius of the core at the narrower side of the taper, $R_2^s$ [$\mu m$] | $3 \times R_1^s$ | Taper shape factor $b_f$ | 1.5 |
| Outer cladding radius at the narrower side of the taper, $R_3^s$ [$\mu m$] | 62.5 | Twist pitch [mm] | 3 (Fig. 4, Fig. 5) |
| Fiber length, L [cm] | 150 | Initially excited mode at the narrower side of the taper | $CV_{0,1}^+$ |

*2.2.2 Twist-induced coupling coefficients*

The next step is to derive CMEs in the presence of twist perturbation, which read (see Supplement 1 for the full derivation)

$$\frac{da_j(z)}{dz} - i\beta_j' a_j(z) = i \sum_q a_q(z) \left( \mathbb{K}_{jq} + \mathbb{Q}_{jq} \right),$$

(12)

where $a_j(z)$ is the amplitude of the $j$-th forward-propagating OAM mode, and $\beta_j'$ represents its propagation constant in the helical frame. Twist induced coupling is characterized by two coefficients. The first, $\mathbb{K}_{jq}$, describes coupling from the $j$-th to the $q$-th forward-propagating OAM mode, arising from perturbations in the permittivity of the medium. The second, $\mathbb{Q}_{jq}$, arises from perturbations in permeability and contributes only when the modes share the same TAM, i.e., $J_j = J_q$ (see Supplement 1). Their explicit form can be expressed in the following compact form:

$$\begin{cases} \mathbb{K}_{jq} = \frac{k_0}{4}\sqrt{\frac{\varepsilon_0}{\mu_0}} \iint \xi_q(r',\varphi').[\delta\varepsilon].\xi_j^*(r',\varphi')\, r'dr'\, d\varphi', & (a), \\ \mathbb{Q}_{jq} = \frac{k_0}{4}\sqrt{\frac{\mu_0}{\varepsilon_0}} \iint \hbar_q(r',\varphi').[\delta\mu].\hbar_j^*(r',\varphi')\, r'dr'\, d\varphi', & (b), \end{cases}$$

(13)

where $\delta\varepsilon$ ($\delta\mu$) are the variation of permittivity (permeability) relative to the ideal fiber defined in Eq. (6). The perturbation tensors are given by $[\delta\varepsilon] = [\varepsilon_r'] - \bar{\varepsilon}_r^*\mathbb{I}$, and $[\delta\mu] = [\mu_r'] - \bar{\mu}_r^*\mathbb{I}$, where $\mathbb{I}$ is the $3 \times 3$ identity matrix.

A closer inspection of the CMEs above, and the explicit forms of the coupling coefficients reveals a key property of twisted fibers. Although the fiber is originally non-magnetic, the coordinate transformation introduces anisotropy in the permeability as well. This activate a magnetic-like modal coupling, which cannot be neglected, as it contains as much information about twisting as permittivity does. In fact, as detailed in Supplement 1, under the zeroth-order vector mode approximation, the magnitudes of the coupling coefficients $\mathbb{K}_{jq}$ and $\mathbb{Q}_{jq}$ are identical, i.e., $|\mathbb{K}_{jq}| = |\mathbb{Q}_{jq}|$. Similar to the case of tapering, these coupling coefficients reveal the existence of a set of selection rules governing modal coupling in twisted RCFs. The rules are given as

$$\begin{cases} \beta_q - \beta_j \pm \{J_q - J_j\}\Upsilon \approx 0, \\ J_q - J_j \mp N = 0, \end{cases}$$

(14)

where $J_q = l_q \pm 1$ and $J_j = l_j \pm 1$ represent the TAM of modes $q$ and $j$, respectively, and $l_q$ and $l_j$ are their corresponding topological charges. Here, $\beta_q$ and $\beta_j$ are the propagation constants of the untwisted modes, i.e., $\Upsilon = 0$, and $N$ denotes the azimuthal order of the perturbation (see Supplement 1). It is important to note that twisted fibers can be fabricated with either right- or left-handed helicity. In this study, without loss of generality, we focus on right-handed structures, for which the second line of Eq. (14) becomes $J_q - J_j = +N$.

Due to the coordinate transformation in Eq. (1), the phase matching condition in twisted fibers resembles that of helical long period gratings of order $J_q - J_j$, with a twist rate of $\Upsilon$ (see Supplement 1). This condition, expressed in the first line of Eq. (14), determines efficient modal coupling, while the TAM selection rule enables interactions among different OAM modes.

To compute the evolution of OAM mode amplitudes in twisted fibers, we begin by solving Eq. (12). This requires first calculating the modal field distributions of the guided modes in an ideal, untwisted fiber, which serve as the basis for evaluating the coupling coefficients defined in Eq. (13). As detailed in Supplement 1, in an ideal rotationally symmetric twisted fiber ($N = 0$), the TAM matching rule (second line of Eq. (14)) reduces to $J_j = J_q$. These modes may satisfy the condition $l_q - l_j = 2$. For generality, we also add a small perturbation to the ideal fiber, accounting for fabrication imperfections, which can cause twisted fibers to behave as onefold-symmetric ($N = 1$) structures [24,25]. This implies that $n^2(r',\varphi')$ becomes $\varphi'$-dependent [26]. Neglecting perturbation terms recovers the ideal case ($N = 0$). Here, we do not specify the explicit form of $n^2(r',\varphi')$ and calculate the coupling coefficients over the entire fiber cross section. While the exact form of the $n^2(r',\varphi')$ can modify the integration domain [14], and the magnitude of the coupling coefficients, the selection rules and physical insights derived here remain unaffected. We then use a Runge-Kutta-based solver to track the evolution of a specific OAM mode as it propagates through the twisted RCF, accounting for all possible twist-induced modal interactions. In Fig. 4(a), we show the twist-induced coupling coefficient, $\mathbb{K}_{jq}$, in a uniform RCF for mode pairs $CV_{0,1}^+$ and $CV_{1,1}^+$, $CV_{1,1}^+$ and $CV_{2,1}^+$, $CV_{2,1}^+$ and $CV_{3,1}^+$, and $CV_{3,1}^+$ and $CV_{4,1}^+$. For comparison, Fig. 4(b) presents the same coefficients in a non-uniform

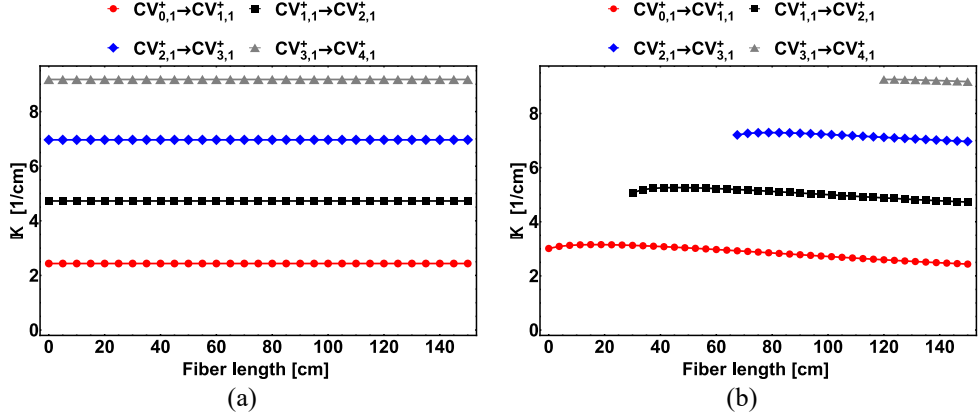

Fig. 4. Twist-induced coupling coefficient $\mathbb{K}$ in (a) a uniform, (b) a non-uniform (tapered) RCF considering the case of $N = 1$. The complete set of simulation parameters used here is given in Table 1

(tapered) RCF. In this case, modal field profiles vary along $z$, making the coupling coefficients $z$-dependent. The tapered fiber supports OAM modes with $l = 0$ and 1 at the narrower side of the taper, and $l = 0, 1, 2, 3$ and 4 at the wider one. Here we only consider the case of $N = 1$. The complete set of simulation parameters used here is provided in Table 1.

A close inspection of Fig. 4(b) reveals that at the narrower side of the tapered fiber, coupling occurs only between modes $CV_{0,1}^+$ and $CV_{1,1}^+$, as these are the only supported modes. As the axial position increases, higher-order modes become guided, enabling further couplings in accordance with the selection rules in Eq. (14).

As shown in Fig. 5(a) and (b), we assess the strength of modal couplings by computing the ratio $\varrho$, as before, substituting $\Omega_{jq}$ with $\mathbb{K}_{jq}$, for all mode pairs presented in Fig. 4(a) and (b). For each pair, the analysis assumes it to be the sole modal interaction present. In comparison with Fig. 3(b), it is evident that taper-induced couplings are negligible compared to the significantly stronger twist-induced couplings. This conclusion is further supported by a direct comparison between Fig. 4(a) and (b), and Fig. 3(a). To finalize the analysis of twist-induced effects, we assess the phase-matching condition in the first line of Eq. (14). Because of the non-uniform nature of tapered fibers, propagation constants vary along $z$. As a result, satisfying the phase-matching condition for OAM modes $q$ and $j$ requires a chirped twisting profile, $\Lambda(z)$, as

$$\Lambda(z) \approx |J_q - J_j| \frac{\lambda}{|n_{eff,q}(z) - n_{eff,j}(z)|}.$$

(15)

However, in the case of uniform fibers, the propagation constants are independent of $z$, and thus Eq. (15) yields a constant twist pitch.

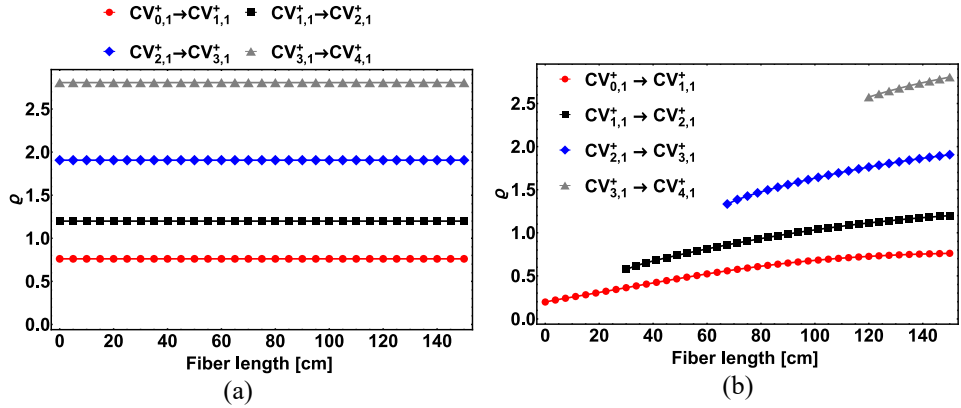

Fig. 5. (a) and (b): The ratio $\varrho$ for all mode pairs presented in Fig. 4(a) and (b), respectively, considering the case of $N = 1$. Each calculation assumes that the corresponding pair is the only interaction present

## 3. Rigorous theoretical framework and simulations

After having discussed the effect of tapering and twisting separately, in this Section we combine these two aspects, providing a comprehensive analysis of OAM mode evolution in twisted, tapered RCFs. Combining Eqs. (8) and (12), and using the slowly varying amplitude approximation, yields the following generalized set of CMEs:

$$\frac{d\tilde{a}_j(z)}{dz} = \sum_q \tilde{a}_q(z) \left\{ i\left(\mathbb{K}_{jq} + \mathbb{Q}_{jq}\right) e^{i(\beta_q - \beta_j \pm \{J_q - J_j\}\Upsilon)z} + \Omega_{jq} e^{i(\beta_q - \beta_j)z} \right\}.$$

(16)

To align with experimentally relevant conditions, we normalize the OAM modes so that the power carried by each mode is given by $P_j = |a_j(z)|^2$. We also assume an adiabatic taper. The twist- and taper-induced coupling coefficients can be calculated using Eqs. (13) and (11), respectively. With these coefficients, Eq. (16) can then be solved numerically to analyze light evolution in twisted, tapered RCFs. To investigate the impact of twist pitch on the output modal content, we consider a twisted, tapered RCF with twist pitches of 3, 2 and 1 mm. For simplicity, the fiber is assumed to support modes with $l = 0$ and 1 at the narrower side of the taper, and $l = 0, 1$ and 2 at the wider end. However, the presented model is general and can accommodate any arbitrary number of modes. All simulation parameters are listed in Table 1. Assuming $CV_{0,1}^+$ is initially excited at the narrower end, Fig. 6(a-c) shows the output power fractions of different modes for the case of $N = 1$. Based on the TAM selection rule, $J_q - J_j = +1$, the only allowed couplings are $CV_{0,1}^+ \to CV_{1,1}^+$ and $CV_{1,1}^+ \to CV_{2,1}^+$.

As seen in Fig. 6(a), the optical power is primarily confined to $CV_{0,1}^+$ and $CV_{1,1}^+$, with only a small fraction in $CV_{2,1}^+$. Reducing the twist pitch to 2 mm and 1mm, panel (b) and (c) in Fig. 6, respectively, significantly increases the power transferred to $CV_{1,1}^+$ and $CV_{2,1}^+$. These results underscore the critical role of twist pitch in controlling modal power distribution in tapered twisted RCFs, demonstrating that fine tuning of the twist pitch can enable coupling from an initially excited mode to a higher-order OAM mode at the output [27]. Also, the modal evolution over the final 4 cm, Fig. 6(d), reveals rapid oscillation of modal powers along z. Nevertheless, analysis of the averaged modal power along the fiber, further confirms the behavior observed in Fig. 6(a-c). This behavior can be attributed to two factors. First, any change in twist pitch will modify the phase-matching condition, first line of Eq. (14), affecting energy transfer between interacting modes. Second, our analysis confirms that twist-induced

coupling coefficients are inversely proportional to the twist pitch, leading to stronger modal

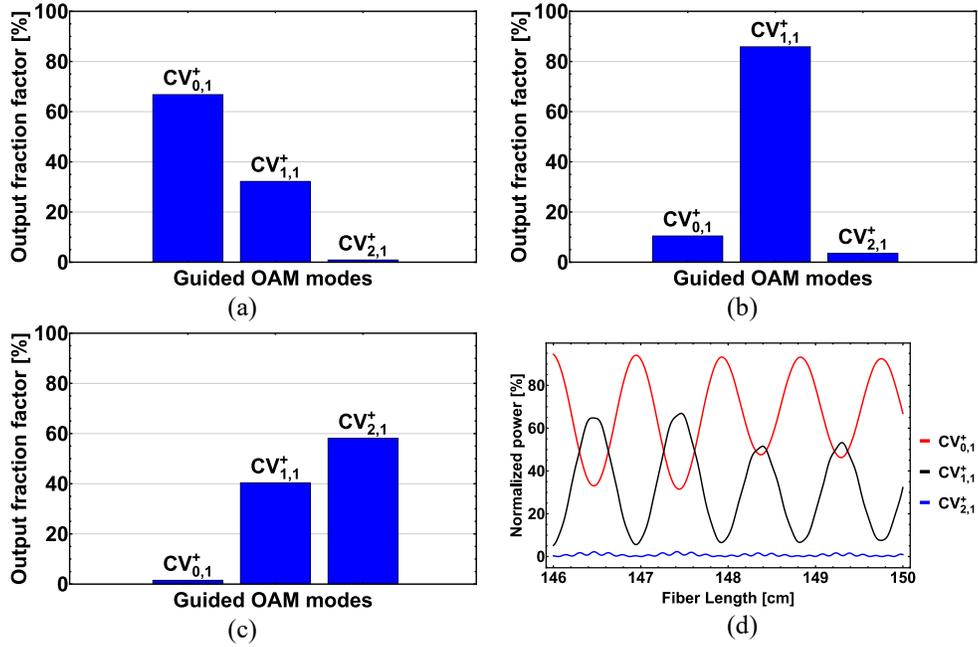

Fig. 6. Output fraction factor of different OAM modes in a twisted tapered RCF with twist pitches of (a) 3 mm, (b) 2 mm, and (c) 1 mm considering $N = 1$. (d) modal evolution of three OAM modes in the same fiber as in panel (a) over the final 4 cm.

interactions as the pitch decreases.

In the discussion regarding the selection rules in Eq. (14), we demonstrated that couplings with $l_q - l_j = 2$ are also possible when interacting modes share the same TAM (the case $N = 0$). This coupling is typically neglected because the mismatch in propagation constants suppresses efficient power transfer. However, for sufficiently long propagation distances or large coupling coefficient, this power transfer can be significant. As a final analysis, we qualitatively assess this effect in Fig. 7(a) and (b) for twist pitches of 3 mm and 1 mm, using

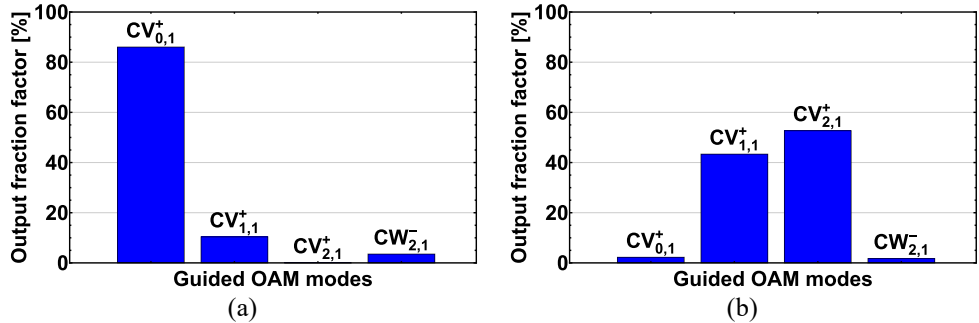

Fig. 7. Output fraction factor of different OAM modes in a twisted tapered RCF with pitches of (a) 3 mm, (b) 1 mm considering both $N = 1$ and $N = 0$.

the same simulation parameters as in Fig. 6.

The presence of this additional coupling term modifies the power distribution among guided modes, although the overall transfer behavior remains similar to Fig. 6. This result demonstrate that for an accurate analysis of OAM modes dynamics in twisted fibers, this coupling should be considered, as it can influence the optimization of twist pitch.

## 4. Conclusion

In this work, we have analyzed the interaction between different OAM modes propagating through twisted tapered RCFs using the formalism of transformation optics and coupled mode theory. We derived the matching conditions that govern the coupling between OAM modes induced by tapering and twisting and evaluated the corresponding coupling coefficients. Our analysis indicates that coupling due to tapering is generally much weaker than that caused by twisting and can therefore be neglected in the calculations. Moreover, we demonstrated that in twisted fibers, perturbations in both permittivity and permeability should be considered. In particular, under the zero-order vector mode approximation, the coupling coefficients due to perturbation in permittivity and permeability are of equal magnitude, highlighting the necessity of accounting for permeability perturbation. Furthermore, based on the rigorous theoretical framework developed, we demonstrated how changes in twist pitch influence the output OAM mode content. The theoretical framework developed here is general and can be extended to consider arbitrary fiber geometries, such as conventional fibers or uniform fibers, as well as cladding modes. This makes it a versatile tool for analyzing the propagation characteristics of twisted optical fibers.

**Funding.** The authors acknowledge the financial support from the Research Council of Finland Flagship Programme (PREIN - decision Grant No. 320165) and the Horizon Europe EIC Pathfinder Open (101096317) project.

**Disclosures.** The authors declare no conflicts of interest.

**Data availability.** Data underlying the results presented in this paper are not publicly available at this time but may be obtained from the authors upon reasonable request.

## Supplemental document

See Supplement 1 for supporting content.